\documentclass[11pt]{article}
\usepackage{graphicx}
[1999/03/29 PSNFSS v.7.2
Times font as default roman
: S Rahtz]

\begin{document}
\renewcommand{\thefootnote}{\fnsymbol{footnote}}
\sloppy
\newcommand{\rp}{\right)}
\newcommand{\lp}{\left(}
\newcommand \be  {\begin{equation}}
\newcommand \ba {\begin{eqnarray}}
\newcommand \ee  {\end{equation}}
\newcommand \ea {\end{eqnarray}}

\title{Andrade and Critical Time-to-Failure Laws in Fiber-Matrix Composites:
Experiments and Model}
\thispagestyle{empty}

\author{H. Nechad$^1$, A. Helmstetter$^2$, R. El Guerjouma$^1$ and D. Sornette$^{2,3,4}$\\
$^1$ GEMPPM Groupe d'Etude de M\'etallurgie Physique et de Physique des Mat\'eriaux,\\
CNRS UMR5510 and INSA de Lyon, 20 Avenue Albert Einstein, 69621 Villeurbanne Cedex, France\\
$^2$ Institute of Geophysics and Planetary Physics, \\
University of California, Los Angeles, California 90095\\
$^3$ Department of Earth and Space Science,\\
University of California, Los Angeles, California 90095\\
$^4$ Laboratoire de Physique de la Mati\`{e}re Condens\'{e}e\\ 
CNRS UMR6622 and Universit\'{e} de Nice-Sophia Antipolis, \\
B.P. 71, Parc Valrose, 06108 Nice Cedex 2, France}

\maketitle

\centerline{\bf Abstract}

We present creep experiments on fiber composite materials. 
Recorded strain rates and acoustic emission (AE) rates 
exhibit both a power law relaxation in the primary creep regime and a 
power-law acceleration before global failure. 
In particular, we observe time-to-failure power laws
in the tertiary regime for acoustic emissions over four decades in time.
We also discover correlations between some characteristics of
the primary creep (exponent of the power-law and duration) and the
time to failure of the samples. 
This result indicates that the tertiary regime is dependent on the
relaxation and damage processes that occur in the primary regime and
suggests a method of prediction of the time to failure based on the early
time recording of the strain rate or AE rate.
We consider a simple model of representative elements, interacting via
democratic load sharing, with a large heterogeneity of strengths. 
Each element consists of a non-linear dashpot in parallel with a spring.
This model recovers the experimental observations of the
strain rate as a function of time.

\subsection*{Keywords: creep, rupture, Andrade law, acoustic emission,
  composites,  time-to-failure singularity, fiber bundle models, shear-thinning rheology, Eyring rheology }

\section{Introduction}

Time-dependent deformation of a material subjected to a constant stress level is known as creep. 
In creep, the stress is below the mechanical strength of the material, so that the rupture does 
not occur upon application of the load. It is by waiting a sufficiently long time that the cumulative 
strain may finally end in catastrophic rupture. Creep is all the more important, the larger the 
applied stress and the higher the temperature. The time to creep rupture is controlled by the 
stress sign and magnitude, temperature and microstructure. 
Since past decades, people have studied the creep rupture phenomena through direct experiments 
(Liu and Ross, 1996; Guarino et al., 2002; Lockner, 1998) as well as through different models 
(Miguel et al., 2002, Ciliberto et al., 2001; Kun et al., 2003; Hidalgo et al., 2002;
Main, 2000; Politi et al., 2002; Pradhan and  Chakrabarti, 2004; Turcotte et al., 2003;
 Shcherbakov and Turcotte, 2003;  Vujosevic and Krajcinovic, 1997; Saichev and Sornette, 2003).
If a lot of works were devoted to homogeneous materials like metals
(Ishikawa et al., 2002) and ceramics (Goretta et al., 2001; Morita and Hiraga, 2002) 
many recent studies concerns heterogeneous materials like composites and rocks 
(Liu and Ross, 1996; Guarino et al., 2002; Lockner, 1998).
The knowledge of the failure properties of composite materials are of great importance because 
of the increasing number of applications for composites in engineering structures. The long-term 
behavior of these materials, especially polymer matrix composites is a critical issue for many 
modern engineering applications such as aerospace, biomedical and civil engineering infrastructure. 
Viscoelastic creep and creep-rupture behaviors are among the critical properties needed to
assess long-term performance of polymer based composite materials. The knowledge of these 
properties is also required to design material microstructures which can be used to 
construct highly reliable components. 
For heterogeneous materials, the underlying microscopic failure 
mechanism of creep rupture is very complex depending on several characteristics of the specific 
types of materials. Beyond the development of analytical and 
numerical models, which predict the damage history in terms of the specific parameters 
of the constituents, another approach is to study the similarity of creep rupture with 
phase transitions phenomena (Andersen et al., 1997).

Creep is often divided into three regimes: (i) the primary creep regime
corresponds to a decay of the strain rate following the application of the
constant stress according to the so-called Andrade's law corresponding
to a power law decay with time (Andrade, 1910); (ii) the secondary regime describes 
a quasi-constant strain rate, which evolves towards the (iii) tertiary creep 
regime in which the strain rate accelerates up to rupture. Andrade's law is similar 
to the power-law relaxation of the seismic activity (``aftershocks'') triggered by the
static stress change induced by a large earthquake, known as Omori's law (Omori, 1894).
A power-law acceleration of seismicity has been reported before several large
individual earthquakes (Sykes and Jaum\'e, 1999; Sammis and Sornette, 2002)
but is not systematically observed.
In contrast with rupture experiments, the acceleration of seismic activity before
an earthquake is observed systematically only when averaging over many sequences.
This average acceleration can be 
explained by the physics of earthquake triggering (Helmstetter et al., 2003). 
Rupture experiments performed at constant stress rate or strain rate
can also produce a power-law acceleration of acoustic emission before
failure (Anifrani et al., 1995; Garcimartin et al., 1997; Guarino et al.,
2002; Johansen and Sornette, 2000). 
This power-law singularity may have a different origin than creep failure.
For instance, critical rupture at constant stress rate occurs in networks 
of purely elastic elements with a heterogeneous distribution of strengths (Daniels, 1945).

The purpose of the present work is to present new experiments on 
well-controlled fiber composite materials and to explain these experiments, 
using a simple model of representative elements in the framework 
of fiber bundles models (Daniels, 1945).
In our model, each element of the system interacts via democratic load sharing, 
and is characterized by a rupture threshold with a large heterogeneity of strengths. 
Our emphasis is in obtaining a complete description of all three regimes at 
the same time, viewed simultaneously from strain measurements as well as non-destructive
 acoustic emission recordings. Our approach 
strives to determine the crucial physical ingredients to capture the complete phenomenology 
observed in the experiments. 
The organization of the paper is the following. Section 2 presents the experimental design. 
Section 3 describes the experimental results for the strain rate and acoustic emission data. 
Section 4 presents our modeling effort, with a series of three models, from the simplest to 
the most relevant for our experiments. Section 5 concludes.

\section{Experimental design}

\subsection{Materials and preparation of the samples} 
\begin{center}
\begin{figure}

\includegraphics[width=0.9\textwidth,angle=-90]{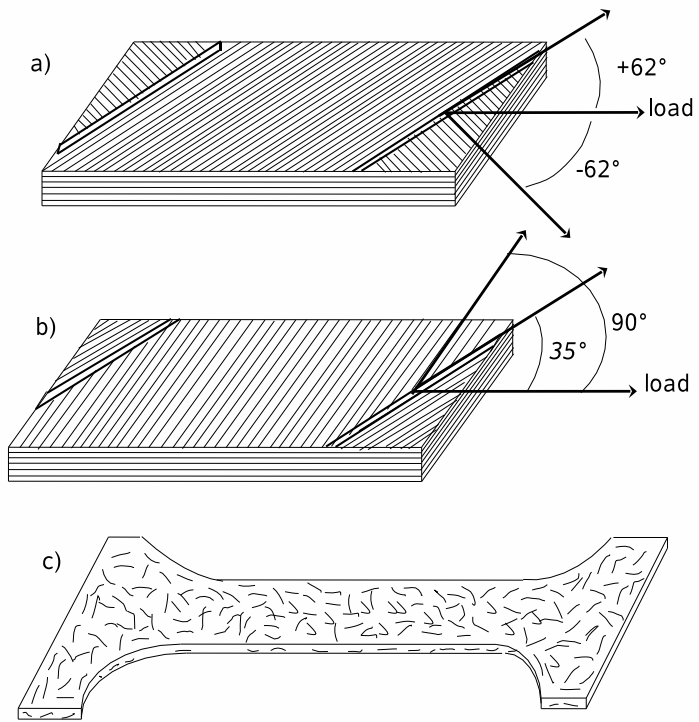}
\caption{\label{crpl}  Structure of Cross angle ply laminates (a)
  $[\pm62^o]_{12}$ and (b) $[90^o/35^o]_{12}$ and (c)  Sheet Molding Compound (SMC) composites.}
\end{figure}
\end{center}

The experimental work is carried out on two types of angle cross ply
glass/polyester composite materials, and on Sheet Molding Compound
(SMC) composites. For all the composites, E glass fibers are used. The
density of the fibers is 2.6 g/cm$^3$. Its tensile strength and modulus
are 3.2 and 76 GPa, respectively. Rectangular plates of angle cross
ply composites are obtained by a hand lay up technique in the
lab. This method consists in applying successively into a mold surface
a release agent, a layer of resin, a layer of
unidirectional reinforcement and to impregnate the reinforcement by
hand by means of a roller. Two cross angle ply laminates denoted
$[\pm62^o]_{12}$ and $[90^o/35^o]_{12}$ are thus fabricated. The angles are measured
with respect to the loading direction as shown in Figure  \ref{crpl}. The
stacking sequence consists of 12 layers. Polymerization is achieved at
room temperature during approximately 12 hours under pressure. The
fiber volume fraction is 75\%. 

The SMC composites are of commercial quality. They consist of a
combination of polyester resin, calcium carbonate filler,
thermoplastic additive and random oriented short glass fibers, in the
form of a sheet. The sheet contains all the components needed for
molding the final part (resin, reinforcement, filler, catalyst, low
profile additives, etc.) in a malleable and non-tacky sheet. Its
characteristics allow it to fill a mould under the conditions of
molding temperature and pressure employed. In our case, the SMC
prepreg is made from glass strands chopped to lengths of 2.5 mm,
sandwiched between two layers of film, onto which the resin paste has
already been applied. The prepreg passes through a compaction system
that ensures complete strand impregnation before being wound into
rolls. These are stored for a few days before molding, to allow the
prepreg to thicken to a moldable viscosity. Good mechanical and
physical properties, together with some advantages in production, make
SMC competitive with metals for automotive applications. As shown in
Figure \ref{crpl}, the relatively low fibre volume fraction, which is in our
case about 30\%, and the uncontrolled filler and reinforcement
distribution during processing lead to a more heterogeneous structure 
compared to the angle cross ply composites, with several kinds of stress 
concentrators such as glass fiber ends and interfaces, calcium carbonate 
filler and voids. The result is that, at quite an early stage, loading 
introduces various forms of local damages.
The SMC samples are in the form of 120 mm barbell with 3 mm tickness. 
The $[\pm62^o]$ and $[90^o/35^o]$ specimens are in parallelepiped form and 
have dimensions $14 \times 120 \times 2$ mm$^3$. All samples are cut using a 
diamond wheel saw to give a suitably smooth surface finish with the
minimum sub-surface damage. 

\subsection{Mechanical testing}

The creep tensile tests were performed using a servo-hydraulic mechanical testing system which is
digitally controlled. The specimens were clamped with serrated wedge 
action grips. Special care was exercised while installing the grips to ensure alignment. 
Constant tensile load was applied to the specimens and the resulting strain was recorded. 
As the failure elongation of the polyester resin at room temperature is close to that of 
the glass fiber, the creep tensile tests were conducted at a higher temperature in order 
to have creep tests with substantial times to rupture for all specimens. After many 
exploratory creep tensile tests, we determined a range of temperatures $T$ and 
stresses $s$ giving manageable time to ruptures. Our results below have been 
obtained with $T=60^o$C and $s =15$ MPa for the $[\pm62^o]_{12}$ specimens and 
with $T=60^o$C and $s=22$ MPa for the $[90^o/35^o]_{12}$ specimens. These temperatures 
are below the glass transition temperature of the resin which is $80^o$C. The
SMC specimens were loaded at $s =48$ MPa at $T=100^o$C.
The stress is increased progressively and reaches a constant value after about 10 seconds
(12 sec for  $[\pm62^o]_{12}$ specimens, 9 sec for $[90^o/35^o]_{12}$, and 17 sec for the SMC).

\subsection{Acoustic emission testing}

\begin{center}
\begin{figure}
\includegraphics[width=1\textwidth]{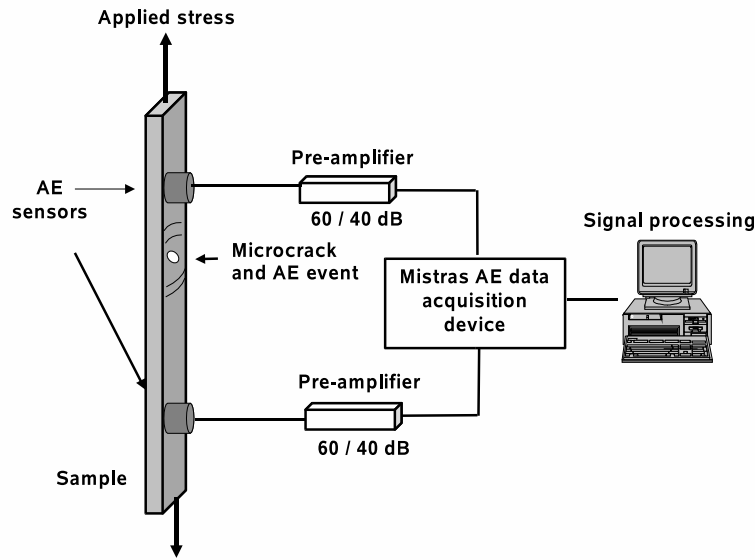}
\caption{\label{aeexp} AE experimental device.}
\end{figure}
\end{center}

An external load applied to the composite 
materials considered in this study results in inhomogeneous stress and strain fields. 
The inhomogeneity of the stress field, coupled with the inhomogeneity of the strength and
fracture properties of the reinforcing elements, the matrix, the interface and the fibers 
lead to a gradual development of damage. For such fiber-reinforced polymer composite laminates,
 failure results from the initiation and coalescence of damages of different types. 
The most typical damage mechanisms  are matrix cracks, fiber matrix debounding, fiber breaks, 
and delaminations. Theoretical and experimental studies of these damage mechanisms in 
these composites are necessary to better understand their ultimate failure and lifetime. 
A number of nondestructive evaluation (NDE) techniques, such as thermography, eddy current, 
radiography, X-ray tomography, ultrasonics, can be used to analyse the damage evolution in 
composite materials. However, the characteristic sizes 
of the matrix cracks, fiber breaks, fiber-matrix disbonds, and ply-damage induced delaminations 
are usually too small to be detected individually by the conventional NDE techniques. 
Acoustic emission (AE) has been proposed to monitor real time damage in composites (Yamaguchi et al., 1991).
In this method, ultrasonic waves generated by the rapid 
release of elastic energy during damage events are detected by piezoelectric sensors.
The corresponding signal is amplified by a pre-amplifier and digitized 
by the AE system, as shown in Figure \ref{aeexp}. 

\begin{center}
\begin{figure}
\includegraphics[width=0.6\textwidth]{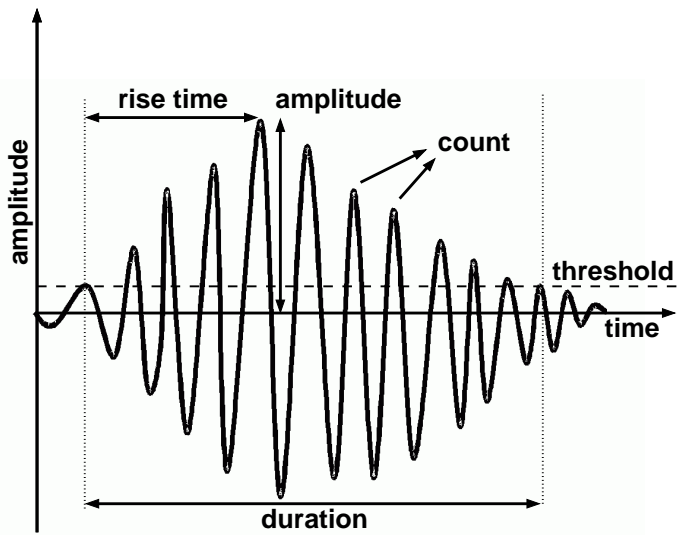}
\caption{\label{aesi}  A typical acoustic emission signal.}
\end{figure}
\end{center}

Generally, the system can work in the parametric or in the transient mode.  
The parametric method is based on the extraction of a number of parameters from individual 
AE signals. A typical AE signal is shown in Figure \ref{aesi}. Some of the AE parameters are 
defined in this figure, including signal amplitude, duration, rise time, decay time, 
thresholds and AE counts. 

Since the early works of Czochralski (1916), Portevin and Le Chatelier (1923),
F\"orster and Scheill (1934), and Kaiser (1950), on metallic materials,  AE analysis 
has been intensely used these last decades to characterize the overall damage accumulation 
in composites (Williams and Reifsnider, 1974; Beattie, 1983;  Yamaguchi et al., 1991;
Bakuckas et al., 1994, Ely and Hill, 1995; Luo et al., 1995; Shiwa et al. 1996).
Williams and Reifsnider (1974) showed that the AE rate generally correlated 
with the rate of stiffness reduction due to damage. Numerous attempts to identify sources 
of the AE signals in composites have been made. Recently, Ono and Huang (1996), 
Prosser et al. (1995), Kloua et al. (1995), and de Groot et al. (1995)
applied the transient waveform analysis for 
AE source recognition. Methods of pattern recognition analysis and neural networks were used 
for the AE signal classifications. It was shown for example by Huguet and al.(2002) for 
unidirectional glass fibre polymer composites that the wave shapes can be associated with particular 
damage mechanisms. These recent results showed that the transient AE analysis method 
based on the full waveform analysis may provide more 
powerful and robust capability to discriminate between the damage mechanisms. 

A hybrid transient-parametric approach to separate overall AE histories into 
the histories for different damage micro-mechanisms in unidirectional composites was 
proposed by Dzenis and Qian (1998). The method was based on the combination of the transient AE 
waveform analysis and multiparameter filtering. The method was recently applied successfully to 
damage evolution analysis of cross-ply and angle-ply composites by Dzenis and Qian (2001).

In the present work, a two channel Mistras 2001 data acquisition system by Physical Acoustics 
Corporation was used for AE recording and analysis. The recorded AE amplitudes range
from $0$ dB to $100$ dB. Two resonant  sensors (200 kHz - 1 MHz) separated by 36 mm were put on the 
specimens using silicon grease as the coupling agent. Preliminary to recording 
during creep experiments, the data 
acquisition system has been calibrated for each sample by using a pencil lead 
break procedure (Nielsen, 1989). In this procedure, a reproducible
acoustic wave is generated in the specimen by breaking a pencil lead
on its surface. 
The lead breakage operation was repeated several times and at different 
locations between the sensors.
The difference in arrival times on the sensors was deduced from 
the first peak of the signals they detected. From this information,
the velocity and attenuation of the AE waves are determined simultaneously.
After this calibration, the AE data 
acquisition was initiated simultaneously with the inception of
mechanical loading. The acoustic emission was 
thus recorded from the beginning of the test all the way to the final failure of the specimen. 
The information on load and strain was continuously obtained from the 
mechanical testing and loading system. 
This information was stored in the parametric AE recording and allowed us
to correlate the AE parameters 
with the load and strain at each time an AE signal was produced.
Signal descriptors such as rise time, counts, energy, duration, amplitude and counts to peak 
were calculated by the software of the MISTRAS system. 
Furthermore, each waveform was digitized and stored. Using the transient analysis, other
parameters were defined, such as the signal waveform, the average frequency, and so on. 
After storage and before processing, the signals were subjected to a linear location procedure 
to determine the location of the AE source. In the analysis of AE data, 
only events located between 
the sensors were used in order to reduce the acoustic noise 
generated by the testing machine and the grips.

We acknowledge that the complexity of wave propagation in composites increases the 
variability of AE signals. Multiple reflections from internal and external
boundaries and the associated mode conversions interfere with the
source wave and change the AE parameters that are detected. 
Despite this, AE can provide a meaningful 
and measurable response to some deformation and damage mechanisms
and/or flow changes. Moreover, our aim in this paper is not to
discriminate between the damage mechanisms but to record the overall
damage accumulation. For this purpose, AE energy events and counts
have shown their relevance (Anifrani et al., 1995;  Garcimartin et al, 1997; 
El Guerjouma et al., 2001).

\section{Creep damage and failure analysis}

\begin{center}
\begin{figure}
\includegraphics[width=0.5\textwidth]{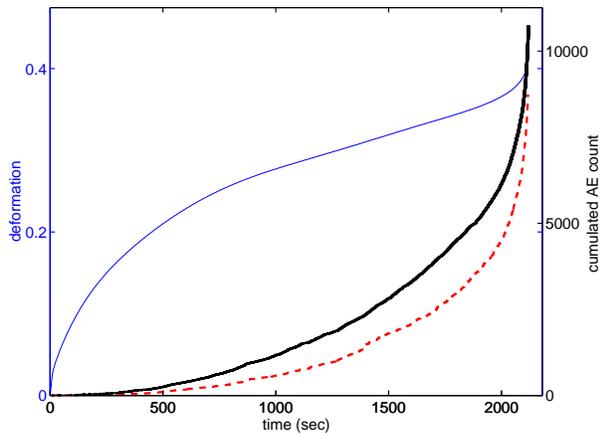}
\caption{\label{eae9035} Creep strain and AE response for $[90^o/35^o]$  angle ply composite \#3.
The thin solid line is the deformation (left axis), the heavy black line is the cumulated 
AE count (right axis) and dashed line is the cumulated energy (arbitrary units).}
\end{figure}
\end{center}

\begin{center}
\begin{figure}
\includegraphics[width=0.5\textwidth]{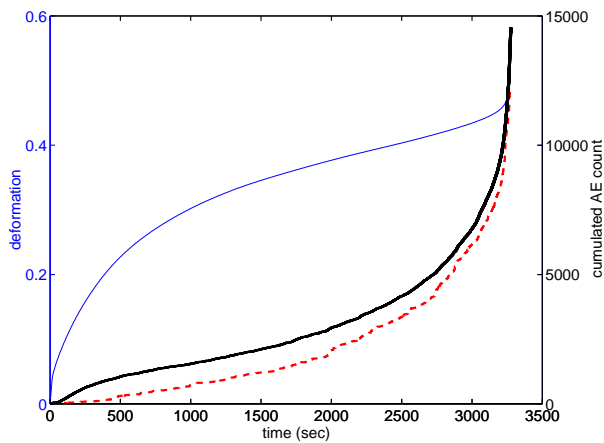}
\caption{\label{eae62} Creep strain and AE response for $[\pm62^o]$ angle ply composite \#4.
Same legend as in Figure \ref{eae9035}.}
\end{figure}
\end{center}

\begin{figure}
\includegraphics[width=0.26\textwidth,angle=-90]{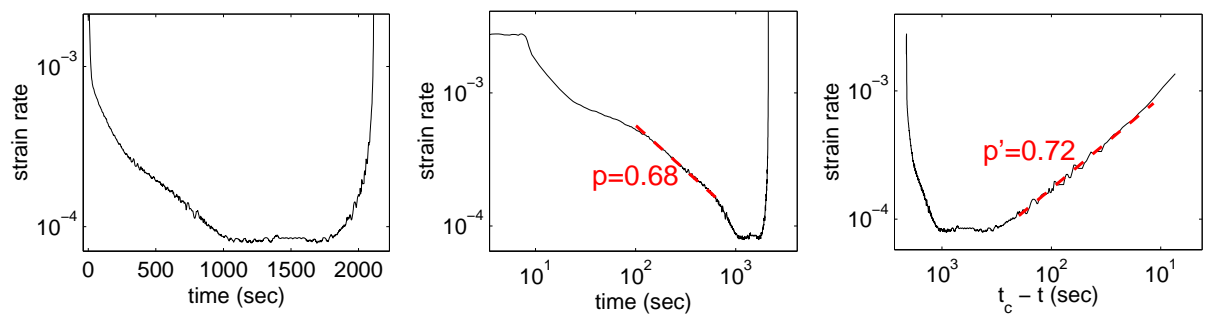}
\caption{\label{dedt9035} Creep strain rate for $[90^o/35^o]$  angle ply composite \#3.
Left panel: full history in linear time scale; middle panel: time is shown in
logarithmic scale to test for the existence of a power law relaxation regime; 
right panel: time is shown in the logarithm of the time to rupture time $t_c$ such
that a time-to-failure power law (\ref{tc1}) is qualified as a straight line. 
}
\end{figure}

\begin{center}
\begin{figure}
\includegraphics[width=0.3\textwidth,angle=-90]{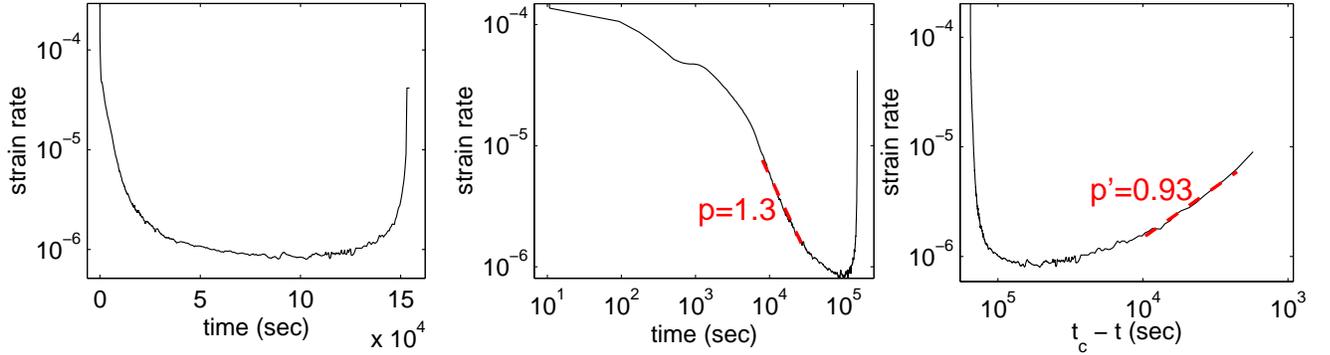}
\caption{\label{dedt62} Creep strain rate for $[\pm62^o]$ angle ply
composite \#4. The three panels are as in figure \ref{dedt9035}.}
\end{figure}
\end{center}

\begin{center}
\begin{figure}
\includegraphics[width=0.26\textwidth,angle=-90]{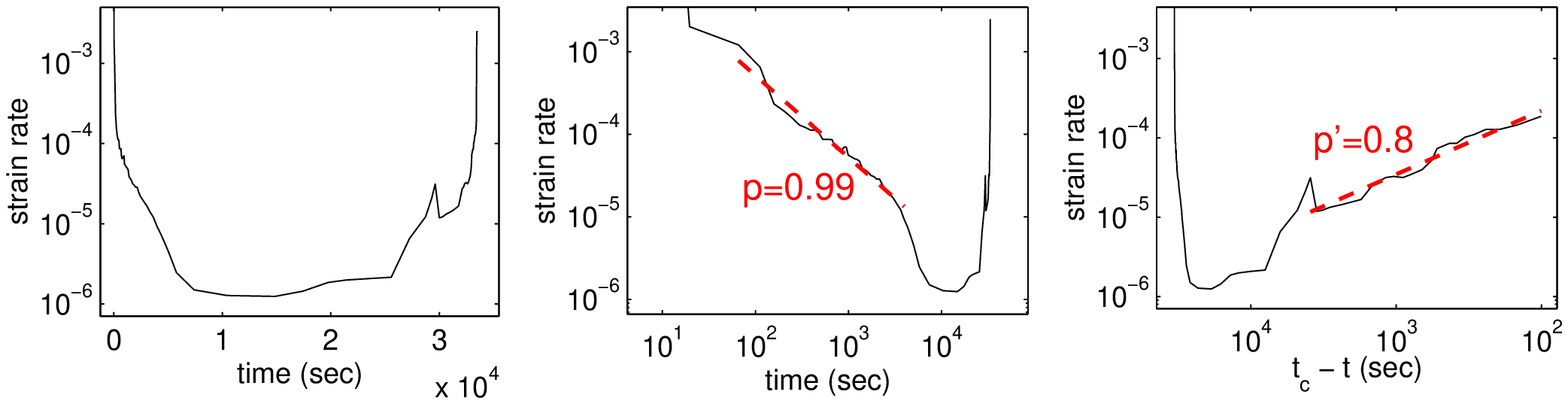}
\caption{\label{dedtSMC} Creep strain rate for SMC composite \#16. 
The three panels are as in figure \ref{dedt9035}.}
\end{figure}
\end{center}

We recorded the strain and the acoustic signals (AE) emitted by microscopic
failure events during the creep tests.
Typical creep responses up to failure of the $[90^o/35^o]$ and $[\pm62^o]$ 
specimens are plotted in Figures \ref{eae9035} and \ref{eae62} respectively 
with the associated AE activity shown as cumulated released energy and counts. 
Normal primary creep transients followed by secondary and tertiary creep were 
observed for almost all samples. Typical strain rate for both composites are 
shown in Figures \ref{dedt9035}, \ref{dedt62} and \ref{dedtSMC}.
Figure \ref{dedtSMC} shows the creep strain rate for the SMC composite \#16. 

These Figures show a rapid and continuous decrease of the strain rate
in the primary creep regime, which can be described by Andrade's law 
\be
{de\over dt} \sim {1 \over t^p}~,
\label{pc1}
\ee
with an exponent $p$ usually smaller than or equal to one (see the middle panels of Figures 
\ref{dedt9035}, \ref{dedt62} and \ref{dedtSMC} and Table \ref{tabdef}).
The crossover for small times is probably due to the fact that the stress
does not reach immediately its constant value, but progressively increases
up to about 10 sec after the start of the experiment.
A quasi-constant strain rate (steady-state or secondary creep) is
observed over an important part 
of the total creep time, and then followed by
an increasing creep rate (tertiary creep regime) culminating in fracture. 
Creep strains at fracture point are large with values around 40\% for angle 
cross ply composites and smaller (around 4\%) for the SMC composites.
The acceleration of the strain rate before failure is well fitted by a
power-law singularity
\be
{de\over dt} \sim {1 \over (t_c - t)^{p'}}~,
\label{tc1}
\ee
with an exponent $p'\leq1$ (see the right panels
of Figures \ref{dedt9035} and \ref{dedt62} and Table
\ref{tabdef}). The critical time $t_c$ determined from the fit of the data
with expression (\ref{tc1}) is generally close
to the observed failure time (within a few seconds). The value of
$t_c$ in Table \ref{tabdef} and Figures \ref{dedt9035} and \ref{dedt62} has been
adjusted to obtain the best fit of the strain rate by equation (\ref{tc1}).

\begin{table}[h]
\caption{Strain rate: values of the critical 
time $t_c$,  determined from the fit of $de/dt$ by a power-law, the
rupture time $t_{end}$, the transition time $t_m$ between the primary and the tertiary 
creep regimes (determined as the time of the minimum of the strain rate), and the values of 
both exponents $p$ defined in (\ref{pc1}) and $p'$ defined in
(\ref{tc1}). The time intervals used to estimate $p$ and $p'$ are given
in column 3 and 5 respectively.}
\label{tabdef}
\vskip 0.2cm
\begin{tabular}{|llclcrlr|}
\hline
sample & $p$ & $[t_{min}-t_{max}]$ & $p'$ & $[t_{min}-t_{max}]$ &  $t_{end}$ & $t_c$ & $t_m$\\
\hline
$[\pm62^o]$ \#1 & 0.84   &    500-3000  &  0.62   &   20-500   &   6617&  $t_{end}$-12 &  4409\\
$[\pm62^o]$ \#2 & 0.79   &     700-8000 &   0.76  &   30-800   &  10214&  $t_{end}$-3  &  7313\\
$[\pm62^o]$ \#3 & 0.99   &     400-2000 &   0.84  &   15-200   &   3267&  $t_{end}$    &  2332\\
$[\pm62^o]$ \#4 & 1.30?  &    8000-30000&   0.93  & 2000-10000 & 154398&  $t_{end}$    & 95478\\
$[\pm62^o]$ \#5 &  0.91  &     500-5000 &   0.75  &   200-3000  &   8517&  $t_{end}+150$ & 4933\\
$[\pm62^o]$ \#6 & 0.97   &     500-3000 &   0.48  &  100-2500  &   6154&  $t_{end}$    &   3436\\
$[\pm62^o]$ \#7 & 1.11   &    1500-7000 &   0.96  &   50-1000  &  13309&  $t_{end}$    &   8313\\
$[90^o/35^o]$ \#1 & 1.04   &     150-1100 &   0.81? &   20-200   &   2307&  $t_{end}$-40 &   1490\\
$[90^o/35^o]$ \#2 & 0.27   &      20-300  &   0.45? &    1-50    &    737&  $t_{end}$-5  &    615\\
$[90^o/35^o]$ \#3 & 0.68   &     100-700  &   0.72  &   10-200    &   2117&  $t_{end}$    &   1214\\
smc \#4	      & 0.74   &      20-100  &   0.96  &    1-20      &    221&  $t_{end}$-1.5&  140.9\\
smc \#6	      & 0.92   &      25-150  &   1.03  &  0.5-80    &    339&  $t_{end}$    &  186.5\\
smc \#10      & 0.90   &      25-100  &    1.07 &  0.5-70   &    238&  $t_{end}$    &  152.2\\
smc \#11      & 0.87   &      25-400  &   0.69  &    1-100    &    686&  $t_{end}$-2  &  411.8 \\
smc \#16      & 0.99   &      25-4000 &   0.85  &  100-4000   &  33539&  $t_{end}$-30 &  14850.\\
\hline
\end{tabular}
\vskip 0.2cm
{\small
`?' : large uncertainty on the values of the exponents, due to a limited range of times used in the 
fitting, or to large fluctuations of the strain rate, or because the results are 
very sensitive to the choice of $t_c$.
}
\end{table}

\begin{figure}
\includegraphics[width=0.3\textwidth,angle=-90]{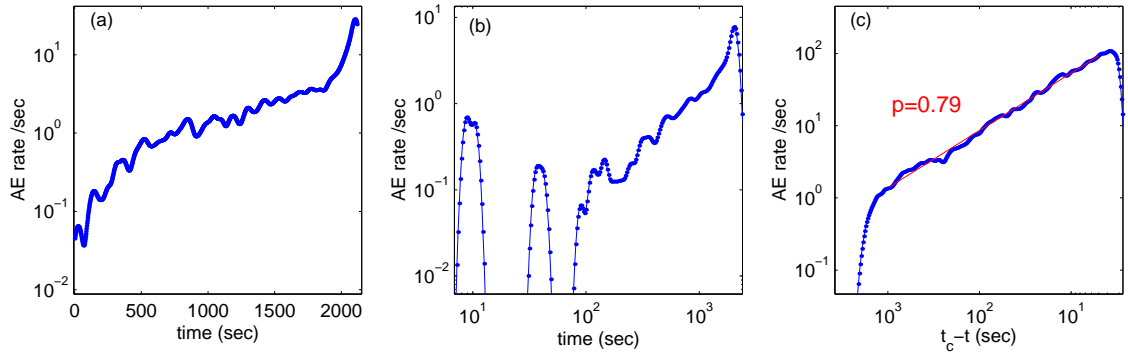}
\caption{\label{EACR9035n3}  Rate of AE events for $[90^o/35^o]$  angle
ply composite \#3. Panel (a): full history in linear time scale; panel (b): time is shown in
logarithmic scale to test for the existence of a power law relaxation regime (for this sample,
this regime is not well-developed); 
panel (c): time is shown in the logarithm of the time to rupture time $t_c$ such
that a time-to-failure power law (\ref{tc1}) is qualified as a straight line. 
}
\end{figure}

\begin{figure}
\includegraphics[width=.3\textwidth, angle=-90]{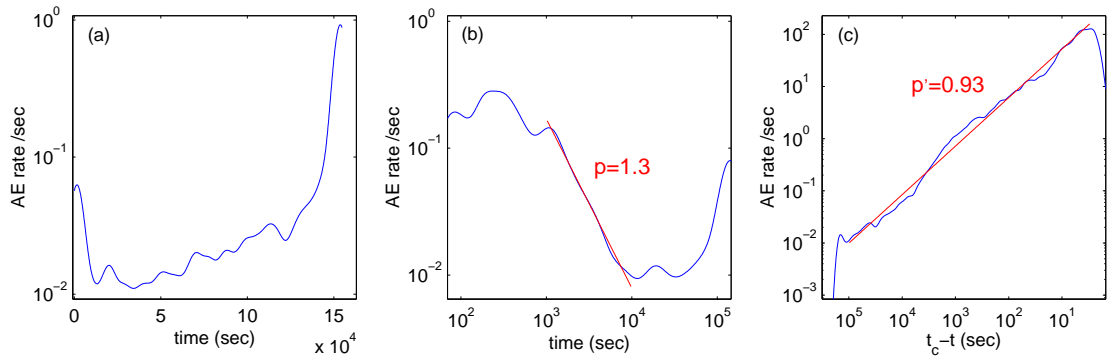}
\caption{\label{EACR62n4} Rate of AE events for $[\pm62^o]$ angle ply
  composite \#4. The three panels are as in figure \ref{EACR9035n3}.}
\end{figure}

\begin{figure}
\includegraphics[width=.3\textwidth, angle=-90]{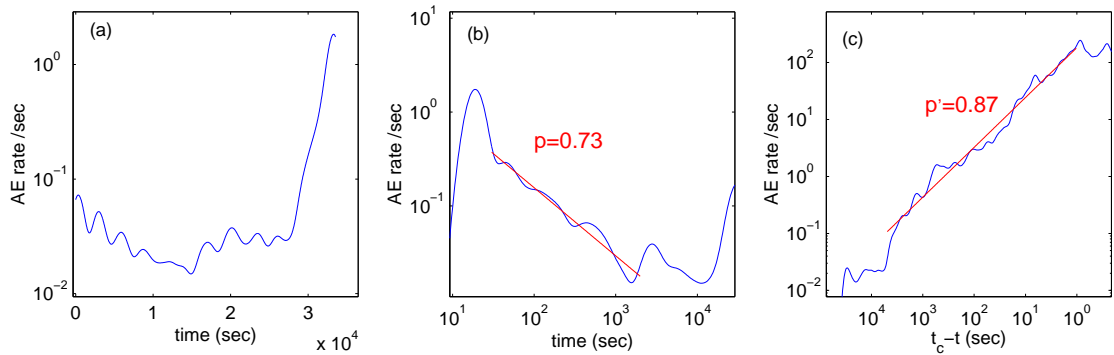}
\caption{\label{EASMCn16} Rate of AE events for SMC  composite \#16. 
The three panels are as in figure \ref{EACR9035n3}.}
\end{figure}

We obtain generally the same temporal evolution for the AE activity as for
the strain rate, as shown in Figures \ref{EACR9035n3}, \ref{EACR62n4}
and \ref{EASMCn16}, except for the $[90^o/35^o]$ samples which do
not display any decrease of the AE activity at early time while they
present a clear relaxation of the strain rate.  
For $[90^o/35^o]$ samples, only one set of fibers is stressed (the 35$^o$ orientation). 
Therefore, the lack of AE activity for the $[90^o/35^o]$ samples may be due to the fact that
the primary AE reflects a reorganization/relaxation between adjacent layers of
 different orientations. We also obtain the same pattern when plotting the AE event 
rate or the rate of AE energy. There are much larger fluctuations for the energy
rate than for the event rate, due to the large distribution of AE
energies, but the crossover time between primary creep and tertiary
creep, and the values of $p$ and $p'$ are similar for the AE event
rate and for the AE energy rate. This suggests that the amplitude 
distribution does not depend on time.


The energy density distributions for all specimens ($[90^o/35^o]$, $[\pm62^o]$
and SMC) are shown in Figure \ref{PA9035}.  
The energy is measured by integrating the square of the amplitude of the
signal over the duration of each event. All materials have
similar energy distributions. 
The distributions of AE energies are approximately power laws with an
exponent of the pdf close to $1.5$. This value is not far from the value $\approx 1.6$
often found for earthquakes in subduction and transform zones (Kagan, 1999;
Pisarenko and Sornette, 2003). 
The roll-off at small energies is due to the detection threshold. 
For the SMC and $[90^o/35^o]$  samples, there is a faster 
than power-law decay for large amplitudes.
Figure \ref{pdfet} shows that the energy distribution does not depend
on time. In particular the largest events do not occur
preferentially close to failure but are also observed during primary
creep, in contrast to one of the predictions of the critical point theory
(Sornette and Vanneste, 1992; Johansen and Sornette, 2000).

\begin{center}
\begin{figure}
\includegraphics[width=0.5\textwidth]{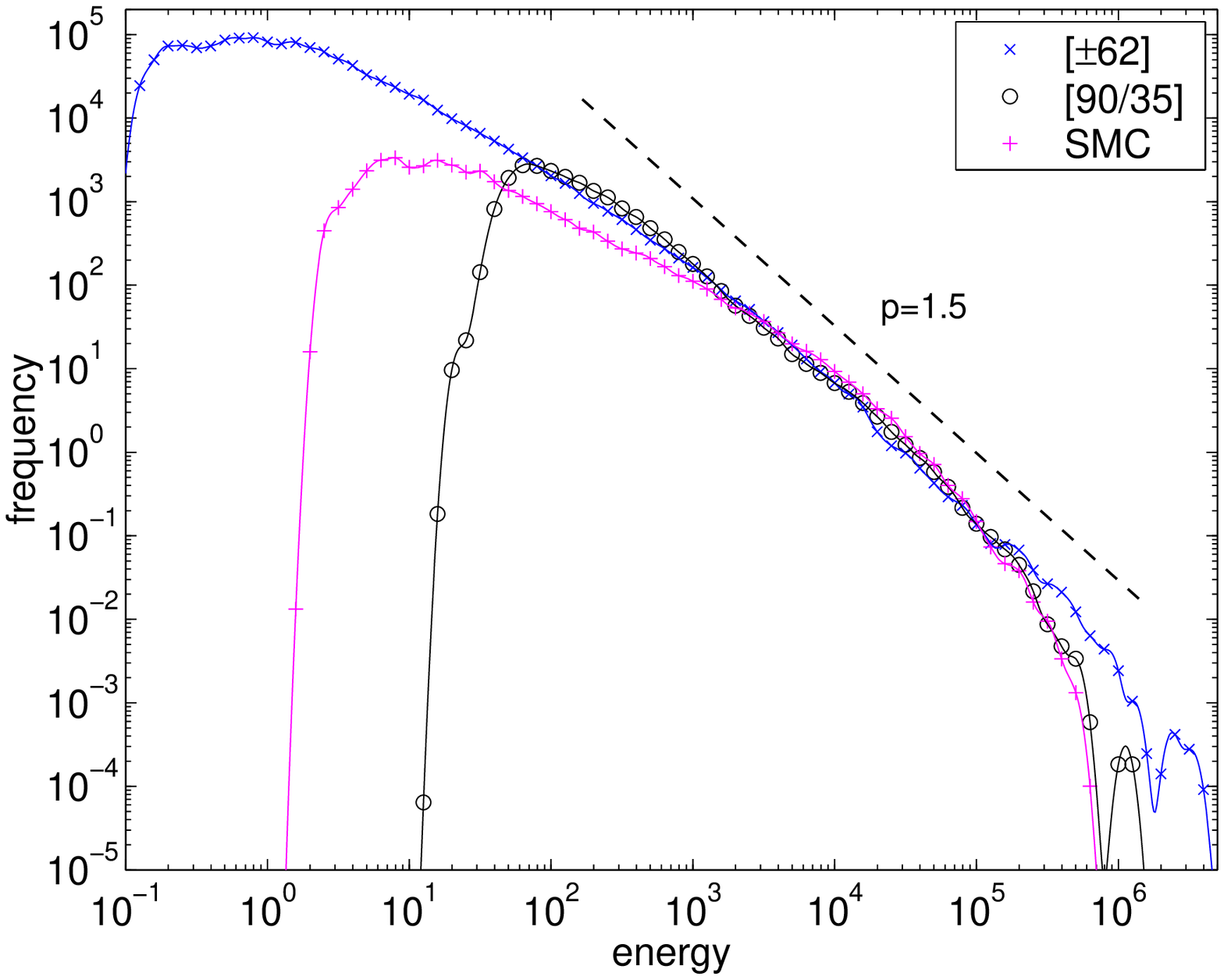}
\caption{\label{PA9035} Energy distribution for a $[90^o/35^o]$, a  $[\pm62^o]$ 
and a SMC specimen.}
\end{figure}

\begin{figure}
\includegraphics[width=0.5\textwidth]{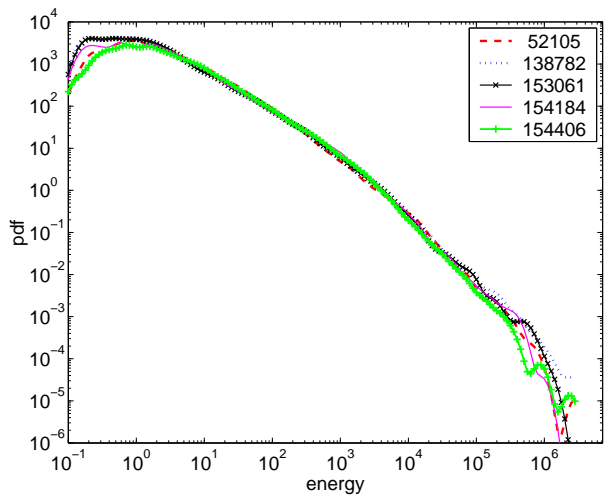}
\caption{\label{pdfet} Energy distribution for the $[\pm62^o]$ 
specimen \#4 at different times, for 5 time windows with 3400 events each. 
The average time (in seconds) of events in each window is given in the caption.}
\end{figure}
\end{center}

We have also analyzed the one-dimensional location
(along the long axis of the samples) of AE events. The AE events are
rather uniformly distributed on the sample, and the spatial distribution 
is almost independent of time. In particular, we do not observe a
significant or systematic localization of AE activity before failure.
 
Tables \ref{tabdef} (strain rates) and \ref{tabae}
(acoustic emissions) give the values of the critical time $t_c$, 
the transition time $t_m$ between the primary and the tertiary creep regimes 
(determined as the position of the minimum of the strain rate), and the values of 
both exponents $p$ defined in (\ref{pc1}) and $p'$ defined in (\ref{tc1}).

\begin{table}[t]
\vskip 0.2cm
\caption{Acoustic emission rate: values of the critical 
time $t_c$, rupture time $t_{end}$, transition time $t_m$, and time
intervals $[t_{min}-t_{max}]$ used to estimate  $p$ defined in
(\ref{pc1}) and $p'$ defined in (\ref{tc1}). $N$ is the number of AE events.}
\label{tabae}
\vskip 0.2cm
\begin{tabular}{|llclcrlr|}
\hline
sample & $p$ & $[t_{min}-t_{max}]$ & $p'$ & $[t_{min}-t_{max}]$ &  $t_{end}$ & $t_c$ & $N$\\
\hline
 $[\pm62^o]$ \#1  &	0.20 & 100-3000  & 0.37  &  1-3000   &	  6624 &   $t_{end}$-9  &   5970\\
 $[\pm62^o]$ \#2  &	0.57  & 100-3000  & 1.06? &  1-300    &	 10226 &   $t_{end}$-12 &   1074\\
 $[\pm62^o]$ \#3  &	0.72  & 100-1000  & 0.91  & 10-3000   &   3273 &   $t_{end}$    &   1453\\
 $[\pm62^o]$ \#4  &	1.30  &1000-10000 & 0.93  &  3-100000 & 154799 &   $t_{end}$-400&  17052\\
 $[\pm62^o]$ \#5  &	0.41  & 100-3000  & 0.48  &0.2-100    &	  8518 &   $t_{end}$    &  11044\\
 $[\pm62^o]$ \#6  &	0.37 & 100-3000  &\multicolumn{2}{c}{no tertiary creep} &6083&&1532\\
 $[\pm62^o]$ \#7  &	0.47  & 500-8000  & 1.13?&   1-100    &	 13417 &   $t_{end}$-106& 2826\\
 $[90^o/35^o]$ \#1 & \multicolumn{2}{c}{no primary creep} &   0.84 & 10-3000&2307 & $t_{end}$-50& 4014\\
 $[90^o/35^o]$ \#2 & \multicolumn{2}{c}{no primary creep} &   0.85 &  2-300 & 737 & $t_{end}$  & 11383\\
$[90^o/35^o]$ \#3 & \multicolumn{2}{c}{no primary creep} &   0.79 &  5-1000&2117 &  $t_{end}$+3 &10724\\
smc \#4       &     0.41? &20-120   &  0.87  &  1-50   &  221 &   $t_{end}$+0.5 &  1979\\
smc \#6       &     1.05? & 15-80   &  0.91  &  1-300  &  339 &   $t_{end}$     &  3406\\
smc \#10      &     1.63? & 15-80	 &  1.10 &  2-100  &  238 &   $t_{end}$-1   &  3058\\
smc \#11      &	0.77   & 25-500	 &  0.94  &  4-100  &  688 &   $t_{end}$-25.1&   546\\
smc \#16      &	0.73   &30-2000  &  0.87  &  1-5000 &33539 &   $t_{end}$     &  7515\\
\hline
\end{tabular}
\vskip 0.2cm
{\small
`?' : large uncertainty on the exponents values, due to a limited range of times used in the 
fitting, or to large fluctuations of the strain rate, or because the results are 
very sensitive to the choice of $t_c$.
}
\end{table}

There is a huge variability of the failure time from one sample to another one, 
for the same applied stress, as shown in Table \ref{tabdef}. 
and in Figure \ref{tmtc}. This implies that one cannot 
predict the time to failure of a sample using an empirical relation between the 
applied stress and the time of failure. There is however another approach suggested 
by Figure \ref{tmtc}, which shows the correlation between the transition 
time $t_m$ (minimum of the strain rate) and the rupture time $t_c \approx t_{end}$:
$t_m$ is found about $2/3$ of the rupture time $t_c$. 
This suggests a way to predict the failure time from the observation of the strain 
rate during the primary and secondary creep regimes, before the acceleration of 
the damage during the tertiary creep regime leading to the rupture of the sample.
As soon as a clear minimum is observed, the value of $t_m$ can be measured and that 
of $t_c$ deduced from the relationship shown in Figure \ref{tmtc}. 
However, there are a few cases where the minima is not well defined, 
such as the one shown in Figure \ref{dedt9035} 
for the $[90^o/35^o]$ angle ply composite \#3 for which the first (smoothed) minimum is
followed by a second similar one. In this case, the application of the relationship
shown in Figure \ref{tmtc} would lead to a pessimistic prediction for the lifetime
of the composite.

\begin{center}
\begin{figure}
\includegraphics[width=0.5\textwidth]{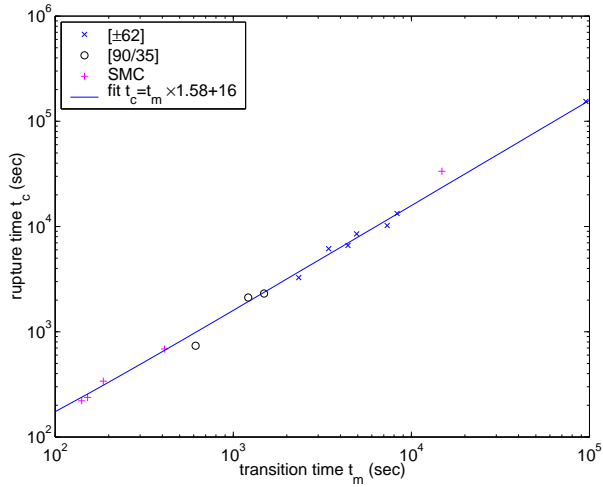}
\caption{\label{tmtc} Relation between the time $t_m$ of the minima of  the 
strain rate and the rupture time $t_c$, for all samples (see table \ref{tabdef}).} 
\end{figure}

\begin{figure}
\includegraphics[width=0.6\textwidth]{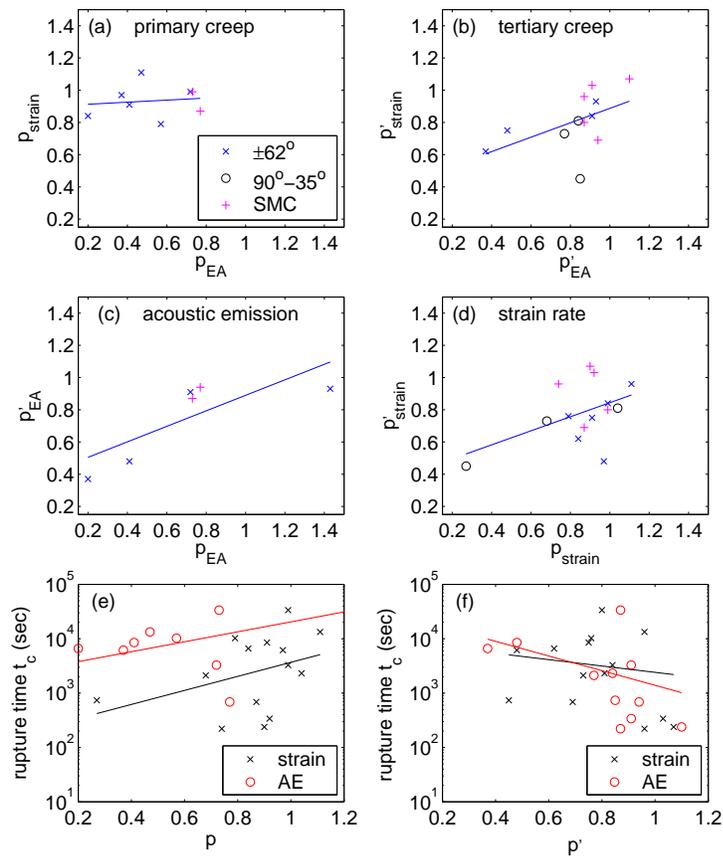}
\caption{\label{cor} Correlation between different characteristics of  the strain rate
and acoustic emission data over all samples (see Tables \ref{tabdef} and \ref{tabae}).
Symbols are data points (see legend in each panel) and solid lines are
linear fits. We removed the data which are the more uncertain (noted
'?' in Tables \ref{tabdef} and \ref{tabae}).}
\end{figure}
\end{center}

Figure \ref{cor} illustrates the correlations between different characteristics
of the evolution of the strain rate and AE activity in the primary creep and 
before failure. There is no correlation between the $p$-value measured using
the strain rate and  the $p$-value of the AE rate (see Figure \ref{cor}a). 
However, due to the rather large uncertainty on the $p$ value 
(which is determined for a more limited time range than $p'$, and for
a smaller number of AE events), it is difficult to ascertain that the 
exponents $p$ of the strain rate and of the AE data are different. 
A positive correlation is found for the exponent $p'$ in the tertiary regime (see
Figure \ref{cor}b) measured for the strain rate and from the AE data.
 There is also a weak positive correlation between the exponent $p$ and $p'$ measured both
for the strain rate (Figure \ref{cor}c) or for the AE rate  (Figure \ref{cor}d).
Finally, we observe a positive correlation between the Andrade exponent $p$ 
and the rupture time, which is more significant for the strain rate
than for the AE rate (Figure \ref{cor}e). 
No such correlation is found between the rupture time and $p'$ (Figure \ref{cor}f). 

The observation that the failure time is correlated with the $p$-value of the 
primary creep suggests that, either a single mechanism is responsible both 
for the decrease of the strain rate during primary creep and for the 
acceleration of the damage during the tertiary creep or, if the mechanisms 
are different nevertheless, the damage that occurs in the primary regime 
impacts on its subsequent evolution in the secondary and tertiary regime, 
and therefore on $t_c$.
It also provides an additional way to predict the failure time from the observation 
of the strain rate or AE activity measured during the primary creep: the smaller 
the $p$-value, the shorter is the time before failure. A similar observation has
been reported in  Kuhn and Mitchell (1993). This correlation between
the $p$-value and $t_c$ comes as an additional information to the
correlation between $t_c$ and the time $t_m$ at the minimum of the strain rate shown in 
Figure \ref{tmtc}. Combining both correlations could help improve the prediction
of the lifetime from the accurate recording of the primary regime and of its 
transition to the tertiary creep.

In contrast, using a fit of the 
AE activity by a power-law to estimate $t_c$ according to formula (\ref{tc1})
works only in the tertiary regime and thus does not exploit the information
contained in the deformation and in the acoustic emissions of the primary and secondary 
regimes which cover 2/3 to 3/4 of the whole history. In practice, one needs
at least one order of magnitude in the time $t_c-t$ to estimate accurately $t_c$ and $p'$,
which means that, if the power-law acceleration regime starts immediately 
when the stress is applied (no primary creep), we cannot predict the rupture time
using a fit of the damage rate by equation (\ref{tc1}) before 90\% of the failure time.
If, as observed in the experiments, the tertiary creep regime starts only at about 
63\% of $t_c$, then we cannot predict the rupture time using a fit of the damage 
rate before 96\% of the failure time. This limitation was one of the motivations
for the development of formulas that extend the description 
of the acoustic emission rates beyond pure power laws 
and that interpolate between the primary and tertiary regimes 
(Johansen and Sornette, 2000; Moura and Yukalov, 2002; Yukalov et al, 2004).
These work however do not impose a relation
between the duration of the primary creep and the failure time
which may be useful to predict $t_c$.

\section{Model construction}

\subsection{Relation with previous works}

Before presenting our model, we briefly discuss previous works that attempted
to explain the power laws for the strain rates
and/or acoustic emission rates observed in the primary and/or tertiary regimes.
Vujosevic and Krajcinovic (1997), Turcotte et al. (2003), Shcherbakov and Turcotte 
(2003) and Pradhan and  Chakrabarti (2004) use systems of elements or fibers
within a probabilistic framework (corresponding to so-called annealed or
thermal disorder) with a hazard rate function
controlling the probability of rupture for a given fiber as
a function of the stress applied to that fiber. 
Turcotte et al. (2003) obtain a finite-time
singularity of the strain rate before failure in fiber bundle models
by postulating a power law dependence of the hazard rate
controlling the probability of rupture for a given fiber as
a function of the stress applied to that fiber.
Shcherbakov and Turcotte (2003) study the same model as Turcotte et al. (2003)
and recover a power-law singularity of the strain rate 
for systems subjected to constant or increasing stresses with an
exponent $p'=4/3$ larger than our experimental results. 
Vujosevic and Krajcinovic (1997) also find this 
power-law acceleration in two-dimensional
simulations of elements and in a mean-field democratic load sharing model, 
using a stochastic hazard rate, but they do not obtain Andrade's law
in the primary creep regime. 
Shcherbakov and Turcotte (2003) obtain  Andrade's law
only in the situation of a system subjected to a constant applied
strain (stable regime). But then, they do not have a global rupture
and they do not obtain the critical power law preceding rupture.
Thus, the models described above do not reproduce at the same
time Andrade's law for the primary regime and a power-law singularity
before failure.

The thermal fuse model (Sornette and Vanneste, 1992; Vanneste and Sornette, 1992) 
introduces a dynamical evolution law for the damage field in systems with long-range 
elastic interactions and frozen heterogeneities. It was initially formulated in the 
framework of electric breakdown. Sornette and Vanneste (1994) latter showed the equivalence
of the thermal fuse-network with  a (scalar) antiplane 
mechanical model of rupture with elastic interaction.
In this model, the damage of each element is a function of the applied stress.
Each element breaks when the damage reaches the rupture threshold.
The global rupture occurs as the culmination of the progressive nucleation, growth and fusion
between microcracks, leading to a fractal network of micro-cracks.
The total rate of damage, as measured for instance by the rate of elastic energy release,
increases on average as $dE/dt \sim 1/(t_c-t)^{p'}$ before failure, with an exponent $p'\geq1$ 
which depends on the damage law.

Miguel et al. (2002) reproduced Andrade's law with $p\approx 2/3$ 
in a numerical model of interacting dislocations, but their model
does not reproduce the tertiary creep regime (no global failure).

Several creep models consider the democratic fiber bundle model (DFBM) 
with thermally activated failures of fibers. 
Pradhan and Chakrabarti (2004) consider the DFBM and add a
probability of failure per unit time for each fiber which depends 
on the amplitude of a thermal noise and on the applied stress.
They compute the failure time as a function of the applied stress and noise 
level but they do not discuss the temporal evolution of the strain rate.
Ciliberto et al. (2001) and Politi et al. (2002)
consider the DFBM in which a random fluctuating force is added on each fiber
to mimic the effect of thermal fluctuations.  Ciliberto et
al. (2001)  show that this
model predicts a characteristic rupture time given by an Arrhenius law
with an effective temperature renormalized (amplified) by the quenched
disorder in the distribution of rupture thresholds. Saichev and Sornette 
(2003) show that this model predicts Andrade's law as well as a power 
law time-to-failure for the rate of fiber rupture with $p=p'=1$, with
logarithm corrections (which may give apparent exponents $p$ and $p'$ smaller than $1$). 

A few other models reproduce both a power-law relaxation in the
primary creep and a finite time singularity in the tertiary regime.
Main (2000) obtains a power-law relaxation (Andrade's law)
followed by a power-law singularity of the strain rate before failure by
superposing two processes of subcritical crack growth, with different 
parameters. A first mechanism with negative feedback dominates in the primary
creep and the other mechanism with positive feedback give the
power-law singularity close to failure. Lockner (1998) gives
an empirical expression for the strain rate as a function of the
applied stress in rocks, which reproduces, among other properties, Andrade's law
with $p=1$ in the primary regime and a finite-time singularity leading
to rupture.

Kun et al. (2003) and Hidalgo et al. (2002) studied numerically 
and analytically a model of visco-elastic fibers, with deterministic dynamics and
 quenched disorder. They considered different ranges of interaction between fibers
(local or democratic load sharing). Kun et al. (2003) derived
the condition for global failure in the system and the evolution of the failure
time as a function of the applied stress in the unstable regime, and analysed
the statistics of inter-event times in numerical simulations of the model.
Hidalgo et al. (2002) derived analytically the expression for the strain 
rate as a function of time.
This model reproduces a power-law singularity of the strain rate before failure 
with $p'=1/2$ in the case of a uniform distribution of strengths (Hidalgo et al., 2002) 
but is not able to explain Andrade's law for the primary creep.
This model gives a power-law decay of the strain rate in the primary creep regime 
only if the stress is  at the critical point, but with an exponent $p=2$ larger
than the experimental values.

Our model is inspired from the work of Kun et al. (2003) and Hidalgo et al. (2002).
We progressively complexify their model to better account for the experimental observations   
and to identify the ingredients necessary to the observation of the 
power-law evolution of the strain rate in the primary and tertiary creep regimes.

\subsection{Model 1: Representative elements with Kelvin rheology}

\subsubsection{Formulation of the model}

We view a composite system as made of a large set of representative elements (RE),
each element comprising many fibers with their interstitial matrix.
Each RE is endowed with a visco-elasto-plastic rheology with 
parameters which may be different from one element to another. The parameters 
characterizing each RE are frozen and do not evolve with time (so-called
quenched disorder). We assume that the applied load is shared
democratically between all RE: each surviving RE is subjected 
to the same stress equal to the total applied force 
divided by the number of surviving RE. This so-called mean-field formulation
allows us to perform analytical calculations and thus to get a better
qualitative understanding of the physics.
This simplifying assumption has been shown to be a good approximation 
of the elastic load sharing for sufficiently
heterogeneous materials (Roux and Hild,  2002;  Reurings and Alava, 2004).

Each RE is modeled as a dashpot of viscous coefficient $b$ put in parallel 
with a purely elastic element of elastic constant $E$, that is, each RE has a 
Kelvin rheology. Applied to our experiments, this would correspond to
a fiber-matrix composite made of long fibers embedded in a visco-elastic matrix
which run all the way through the sample from one extremity to the other.
A given RE is assumed to fail when its elongation/deformation 
$e$ reaches a threshold. The rupture thresholds are distributed according to a cumulative 
distribution $P(e)$, giving the number of thresholds smaller than $e$.
A rupture threshold is uniquely assigned to a given RE and remains constant
throughout the deformation. This first model has been introduced in the language of the 
democratic fiber bundle model (DFBM) extended
to allow for the presence of creep (Kun et al., 2003; Hidalgo et al., 2002), 
with different choices of rupture thresholds.  

We assume that the fraction of broken RE is governed by 
\be
P(e)=1-\left({e_{01} \over  e +e_{02}} \right) ^{\mu} ~, 
\ee
where $e_{01}$ and $e_{02}$ are two constants with $e_{01} \leq
e_{02}$. The fraction $1-(e_{01}/e_{02})^{\mu}$ breaks as
soon as the stress is applied, in agreement with experiments which show
that acoustic emission starts immediately when the stress is applied.
The power-law distribution of rupture thresholds for large $e$ is  
suggested by the large distribution of failure times for the same applied stress
(see Figure \ref{tmtc}). The exponent $\mu>0$ controls the amplitude of the 
frozen heterogeneity of the RE strengths. The larger $\mu$ is, the less 
heterogeneous is the system. We shall consider that $\mu$ is in general larger than $1$
so that the average strength of a RE has a meaning but $\mu$ may be smaller than $2$
for which the variance of the fiber strength is not defined.

The initial stress per RE, before any damage has occurred, is equal to the 
applied load $s$. The equation controlling the deformation $e(t)$ in each surviving
RE as a function of time is 
\be
b {de \over dt} + E e = {s \over 1-P(e)} 
= s ~\left({e+e_{02} \over e_{01}} \right)^{\mu}~~,
\label{jgjklwe}
\ee   
with the initial condition $e(t=0)=0$.
The right-hand-side of this equation gives the stress per remaining RE when the 
deformation (of both the elastic and anelastic parts) is $e$. 
$1-P(e)$ is the fraction of unbroken RE. By construction of the model, 
the RE are associated in parallel and they
all have the same deformation $e$ obeying the same equation (\ref{jgjklwe}).  
The left-hand-side of (\ref{jgjklwe}) expresses the stress within 
a RE as being shared between the dashpot and the elastic element in the Kelvin rheology.

\subsubsection{Condition for global failure \label{condfail}}

\begin{center}
\begin{figure}
\includegraphics[width=0.5\textwidth]{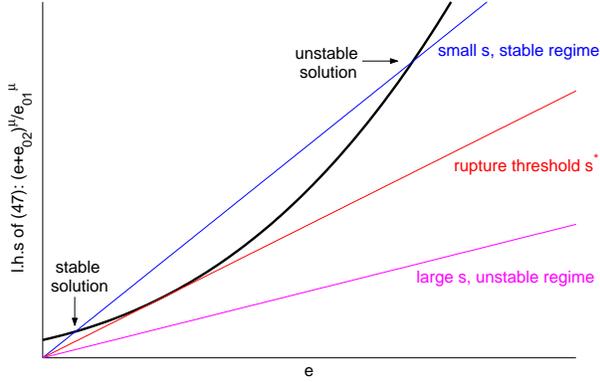}
\caption{\label{stablesol} The thick curve is the left-hand-side
of condition (\ref{wqerwe}) as a function of $e$. The three
thin straight lines are the right-hand-side ${E \over s} e$ 
of (\ref{wqerwe}) for three values of the applied stress $s$. 
The rupture of the system occurs for sufficiently large stress
$s$ for which condition (\ref{wqerwe}) does not have a solution.}
\end{figure}
\end{center}

The system defined by  (\ref{jgjklwe}) has a stable and an unstable regime 
as a function of the applied stress $s$. The system is stable if the differential 
equation (\ref{jgjklwe}) has a stationary solution $de/dt=0$ with $e>0$, i.e. 
if the following equation has a non-trivial solution:
\be
\left({e+e_{02} \over e_{01}}\right) ^{\mu} = {E \over s} e~.
\label{wqerwe}
\ee
For small $s$, the differential equation evolves toward a constant
deformation $e$ given by the smallest of the two solutions of (\ref{wqerwe})
which is stable, as shown in Figure \ref{stablesol}. There is a threshold
$s^*$ above which eq.~(\ref{wqerwe})
does not have a solution and ${de \over dt}$ grows without bound
leading to the global failure of the system.

\subsubsection{Properties of the model}

At short times for which the deformation is small, we have $e \ll e_{02}$.
In this regime $(e+e_{02})^{\mu} \approx {e_{02}}^{\mu} (1 + \mu e /e_{02})$.
We can thus rewrite (\ref{jgjklwe}) as
\be
b {de \over dt} + E e = s ~\left({e_{02} \over e_{01}} \right)^{\mu}~
\left(1 + \mu {e \over e_{02}}\right)~,
\label{jgjkslwssesss}
\ee
whose solution is (with initial condition $e(0)=0$)
\be
{de \over dt} = {s \over b}~\left({e_{02}\over e_{01}}\right)^{\mu} ~
e^{ -t \left ( {E\over b}  - {s\mu \over b e_{02}} 
\left({e_{02}\over e_{01}}\right)^{\mu} \right) }~.
\label{mgvmle}
\ee
After a jump at $t=0$, the deformation rate decays exponentially
with time if the stress is small, or starts to accelerate at $t=0$
without a primary creep regime if the stress or $e_{02} / e_{01}$ are 
large enough so that the term in the exponential of (\ref{mgvmle}) is positive.

In the unstable regime $s>s^*$, the deformation rate accelerates before failure.
Close to failure, we have $e \gg e_{02}$ and we can rewrite
(\ref{jgjklwe}) as
\be
b {de \over dt} + E e = s ~\left( {e \over e_{01}}\right)^{\mu}~.
\label{jgjkslwsse}
\ee   
As the deformation increases, eventually the term in the right-hand-side
of (\ref{jgjkslwsse}) dominates over $E e$, which yields the asymptotic
equation
\be
b {de \over dt}  = s ~\left( {e \over e_{01}}\right)^{\mu}~.
\label{jgjkslwsssse}
\ee
Its solution is
\be
e(t) \approx \left( {b {e_{01}}^{\mu} \over s(\mu -1)}\right)^m~{1 \over (t_c-t)^m}~,~~
{\rm where }~~m = 1/(\mu -1)~.
\label{mvgmle}
\ee
The cascades of RE ruptures give rise to a finite-time singularity, i.e., the
deformation diverges in finite time. The critical time $t_c$ is fixed by matching this
asymptotic regime (\ref{jgjkslwsssse}) with the short-time solution (\ref{mgvmle}).
Of course, in reality, there are no divergences in a system of finite size. 
This solution only expresses a very fast
acceleration of the deformation leading to a failure of the system in finite time at or very
close to $t_c$. The deformation rate is thus proportional to 
\be 
{de \over dt}  \sim {1 \over (t_c-t)^{\mu/(\mu-1)} }~.
\label{mvgmle22}
\ee
Since $\mu >1$, the exponent $p'=\mu/(\mu-1)$ 
is larger than $1$, in contradiction with the experimental results
shown in Tables \ref{tabdef} and \ref{tabae}.

\begin{figure}
\includegraphics[width=0.35\textwidth, angle=-90]{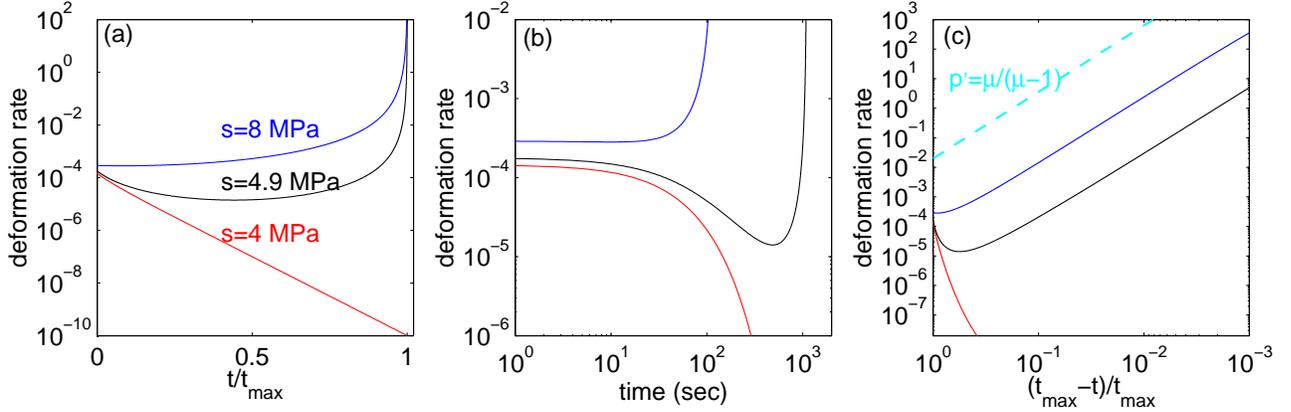}
\caption{\label{mod1} Strain rate $de/dt$ estimated by the
numerical solution of eq. (\ref{jgjklwe}) for different values of the
stress $s$, and with parameters $E=20$ GPa, $\mu=1.8$, $e_{01}=0.003$,
$e_{02}=0.015$, $b=5\times10^{11}$ Pa.sec. 
Panel (b) illustrates the absence of Andrade's law in the 
primary regime.
Panel (c) illustrates the power law acceleration of the 
strain rate before failure for $s=4.9$ MPa and $s=8$ MPa with an 
exponent asymptotically equal to $p'=\mu/(\mu-1)=2.25$ (dashed line). 
The time $t_{max}$ used to normalize the curves
in (a) and (c) is the rupture time (for $s=4.9$ MPa and $s=9$ MPa) or
the time when $de/dt$ decreases below $10^{-14}$ (numerical accuracy) for 
$s=4$ MPa.}
\end{figure}

Figure \ref{mod1} shows the numerical solution of the differential equation 
(\ref{jgjklwe}) for different values of the stress, which are in good
agreement with the approximate solutions (\ref{mgvmle}) for the primary creep
and (\ref{mvgmle22}) close to failure.

Our experiments record not only the deformation rate $de/dt$ 
but also the rate of acoustic emissions
(number of counts per unit time) radiated by the sample. 
If we assume that the rate of acoustic emissions is proportional
to the failure rate of RE, the acoustic emission rate close to failure
is given by
\be
{dN \over dt} = {dP \over de}~{de \over dt} \sim {1 \over (t_c-t)^{1-m\mu}} \sim 
(t_c-t)^{1/(\mu-1)}~,
\label{gmels}
\ee
where we have used (\ref{mvgmle}) to get the last expression on the right.
Thus, rather than increasing in the final stage of deformation as often observed
experimentally, the acoustic emission count rate goes to zero according to Model 1.
Experimentally, we have observed the same temporal evolution for the
strain rate $de/dt$ and for the AE rate $dN/dt$, in disagreement with
the result (\ref{gmels}).
This simple model may be inadequate to
reproduce AE data. In contrast with the DFBM model (network of elastic
fibers loaded with increasing stress), there is no avalanches in this
model: the rupture of one element does not immediately trigger the rupture of
other elements (Kun et al., 2003). Therefore in our further modelling
work we will focus on the analysis of the strain rate.

In conclusion, while Model 1 exhibits some of the
qualitative properties reported in our experiments
(initial decay of the deformation rate and power law acceleration of the 
deformation rate leading to rupture), it
fails to reproduce Andrade's law for the initial decay 
(exponential versus power law) and the 
observation that the exponent $p'$ of the critical acceleration of the
deformation rate is smaller than $1$. 
Note that this model with a uniform distribution of failure thresholds 
gives an exponential relaxation for $s \neq s^*$ and the power law $de/dt \sim 
1/t^2$ only for the critical value $s=s^*$ (Hidalgo et al., 2002).

\subsection{Model 2: Power-law shear-thinning creep with Kelvin fibers \label{secref}}

Let us therefore consider the following modification of the model. Rather than taking
a Newtonian viscous dashpot, we propose to consider a dashpot 
with shear-thinning rheology (Astarita and  Marrucci, 1974)
which consists in replacing $b$ by $b_0 |de / dt|^{n-1}$, with $n<1$.
This shear-thinning power law rheology represents at the mesoscopic RE level
the fiber-matrix interaction which involves micro-cracking and delamination. 
The governing equation replacing expression (\ref{jgjklwe}) reads
\be
b_0 |de / dt|^{n-1} {de \over dt} + E e = {s \over 1-P(e)}~.
\label{mgmssssll}
\ee 

At short times when $e\ll e_{02}$, expression  (\ref{mgmssssll}) reduces to
\be
b_0 |de / dt|^{n-1} {de \over dt} + E e =  s ~\left({e_{02} \over e_{01}} \right)^{\mu}~
(1 + \mu {e \over e_{02}})
\label{mgmll}
\ee 
whose solution for the strain rate is
\be
{de \over dt} = { \left(e_{02} b n g\right)^{1/(1-n)} \over
\left[ e_{02} b n + t g (1-n) (E e_{02} - s f \mu) \right]^{1/(1-n)}}
\label{dedtsol3}
\ee
where the constants $f$ and $g$ are defined by
\ba
f &=&\left({e_{02} \over e_{01}} \right)^{\mu} \\
g &=& \left(s f \over b \right)^{(1-n)/n}
\ea
Expression (\ref{dedtsol3}) describes a constant strain rate at times
smaller than $t^*$ given by
\be
t^*= {e_{02} b n \over g (1-n) (E e_{02} - s f \mu)}
\label{tstar}
\ee
followed by a power law decrease $\sim 1/t^p$ at times $t\gg t^*$ with an
exponent $p=1/(1-n)$ larger than one if $n<1$.

Note that the exponent $p$ does not depend on the exponent of the
distribution of rupture thresholds. Changing the distribution of
rupture thresholds, e.g. by taking an exponential distribution or
assuming that there is no rupture of RE for $e<e_{01}$, does not 
change qualitatively the behavior of the strain rate in the primary 
creep regime. It only changes the crossover
time $t^*$ and the amplitude of the strain rate.
This expression for a power-law rheology
with shear-thinning rheology improves on the exponential relaxation (\ref{mgvmle})
but predicts an exponent $p$ still larger than $1$.

At larger times and for $s > s^*$, the strain rate accelerates up to failure.
When $e\gg e_{02}$, the equation governing the evolution of $e$ becomes
\be
b_0 |de / dt|^{n-1} {de \over dt} + E e = 
s ~\left( {e \over e_{01}}\right)^{\mu}~.
\label{mgmlsdssdl}
\ee
As with (\ref{jgjkslwsse}), when the deformation increases, eventually the term 
in the right-hand-side
of (\ref{mgmlsdssdl}) dominates over $E e$, which leads to the asymptotic equation
\be
b_0 |de / dt|^{n-1} {de \over dt} \approx s ~\left( 
{e \over e_{01}}\right)^{\mu}~,
\label{mgssmlsdssdl}
\ee
whose solution for $de/dt$ reads
\be
{de \over dt} \sim  {1 \over (t_c-t)^{p'}}~,~~ {\rm where }~~p' = {\mu/n  \over (\mu/n) -1}~.
\label{mvgmle2}
\ee
The deformation rate has a power-law singularity with an exponent $p'$ larger than $1$.

\begin{figure}
\includegraphics[width=0.35\textwidth, angle=-90]{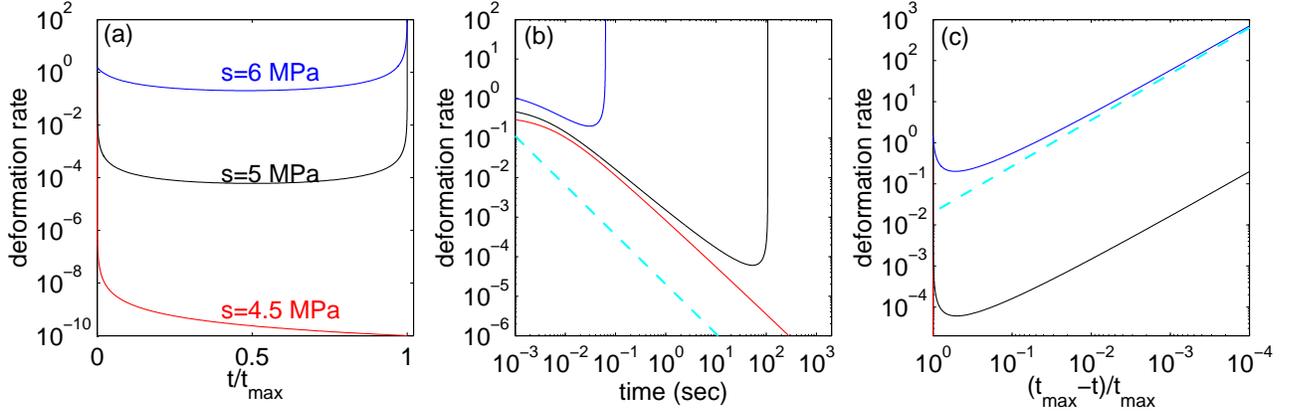}
\caption{\label{mod2} Strain rate $de/dt$ estimated by the
numerical solution of eq. (\ref{mgmssssll}) for different values of the
stress $s$, and with parameters $E=20$ GPa, $\mu=1.8$, $n=0.2$, 
$e_{01}=0.003$, $e_{02}=0.015$, $b=1\times10^{8}$ Pa.sec$^n$. 
Panel (b) illustrates Andrade's law in the 
primary regime with an exponent close to the prediction 
$p=1/(1-n)=1.25$ (dashed line).
Panel (c) illustrates the power law acceleration of the 
strain rate before failure for $s=5$ MPa and $s=6$ MPa with an 
exponent asymptotically equal to $p'=(\mu/n)/[(\mu/n)-1]=1.125$ 
(dashed line). The time $t_{max}$ used to normalize the curves
in (a) and (c) is the rupture time (for $s=5$ MPa and $s=6$ MPa) or
the time when $de/dt$ decreases below $10^{-14}$ (numerical accuracy) for 
$s=4.5$ MPa.}
\end{figure}

Figure \ref{mod2} shows the numerical solution of the differential equation 
(\ref{mgmssssll}) for different values of the stress, which are in good
agreement with the approximate solutions (\ref{dedtsol3}) for the primary creep
and (\ref{mvgmle2}) close to failure.

The shear-thinning power law rheology for the dashpot 
with the Kelvin topology captures the main properties of the
experiments, the power-law evolution of the strain rate in the
primary and tertiary creep regimes. However the quantitative values of $p$
and $p'$ are larger than one in the model, in disagreement with the
values measured in our experiments.

\subsection{Model 3: Eyring rheology} 

We propose here another model which consist of an Eyring dashpot in
parallel with a linear spring of stiffness $E$. The Eyring rheology 
(\ref{eyringdash}) is standard for fiber composites (Agbossou et al., 1995; 
Liu and Ross, 1996). It consists, at the microscopic level,
in adapting to the matrix rheology the theory of reaction rates
describing processes activated by crossing potential barriers.

The deformation $e$ of the Eyring dashpot is now governed by
\be
{de \over dt} = K \sinh (\beta s_1) ~.  \label{eyringdash} 
\ee
where the stress $s_1$ in the dashpot is given as previously by 
$s_1= s /(1-P(e)) -Ee$.

Using the same distribution of rupture threshold $P(e)$ as previously, the 
differential equation for $e$ (with initial condition $e(0)=0$) is thus
\be
{de \over dt} = K \sinh \left( {\beta s \over {e_{01}}^\mu}
(e+e_{02})^{\mu} -\beta E e \right)~.
\label{dif1}
\ee
Let us give the approximative solution of the deformation in the primary 
creep regime for $\beta s_1 \gg 1$ and $e \ll e_{02}$.
In this regime, we have $\sinh(\beta s_1) \approx \exp(\beta s_1)/2$
and $(e+e_{02})^{\mu} \approx {e_{02}}^{\mu} (1 + \mu e /e_{02})$.
We can thus rewrite (\ref{dif1}) as
\be
{de \over dt} \approx {K\over 2} \exp \left( {\beta s {e_{02}}^{\mu} 
\over {e_{01}}^\mu} (1+\mu {e \over e_{02}}) -\beta E e \right)~,
\label{pcc}
\ee
which has the solution
\be
{de \over dt} = {K e_{02} \over 2 e_{02} \exp \left( -\beta s
\left({e_{02} \over e_{01}} \right)^\mu \right) + t K \beta \left(
E e_{02}-\mu s \left({e_{02} \over e_{01}} \right)^\mu \right) }~.
\label{pc}
\ee
Expression (\ref{pc}) predicts that, if the stress is not too large so that 
the term in the exponential of (\ref{pcc}) is negative, ${de/ dt}$ is of the 
Andrade form $\sim t^{-p}$, with an exponent $p=1$ at early times.
For larger $s$, the strain rate starts to accelerate when the load is applied.
Note that the observation of Andrade's power-law creep in this model,
as for the previous models, is not dependent on the choice of the
distribution of rupture thresholds $P(e)$. We obtain for instance the
same behavior  $de/dt \sim 1/(t+c)$ if we take $P(e)=0$ for all $e$
(no broken RE).

An approximate analytical solution of (\ref{dif1}) in the tertiary creep is obtained 
by neglecting $e_{02}$ compared with $e$. Close to failure, for large $e$, the linear 
term $E e$ is small compared with $s_2={ s \over {e_{01}}^\mu} (e+e_{02})^{\mu}$ if $\mu>1$.
This leads to the equation
\be
{de \over dt} \approx {K \over 2} \exp \left( {\beta s {e}^{\mu} \over {e_{01}}^\mu} \right)~.
\label{dif1ss}
\ee
Its solution is to leading logarithmic order (to be defined just below)
\ba
e(t) &=& A ~\left[ -\ln(t_c-t)\right]^{1 \over \mu} ~,  \label{mmwl}  \\
{d e \over dt} &=& {A \over \mu} ~\left[ -\ln(t_c-t)\right]^{{1 \over \mu}-1}~
{1 \over t_c -t}~, \label{mgmss}
\ea
where  $A$ is given by
\be
A=e_{01} \left(\beta s \right)^{-1/\mu}~.
\ee
The solutions (\ref{mmwl}) and (\ref{mgmss}) are correct up to 
$\ln(\ln(t_c-t))$ correction terms for $e(t)$ and up to $\ln(t_c-t)$ terms
for ${d e \over dt}$. This means that only the leading power law $1/(t_c -t)$
is obtained reliably in the self-consistent solution (\ref{mgmss}), while 
logarithm corrections are not determined. 
The solutions (\ref{pc}) and (\ref{mgmss}) are of the same form with $p=p'=1$
as the Langevin-type model solved by Saichev and Sornette (2003). 
This may not be surprising since the Eyring rheology describes, at the microscopic level,
processes activated by crossing potential barriers, which are explicitly
accounted for in the thermal fluctuation force model (Saichev and Sornette, 2003).

Expression (\ref{dif1}) shows, similarly to the discussion in section \ref{condfail} 
on the Condition for global failure for model 1
with equation (\ref{wqerwe}), that rupture will occur only if the argument 
$B \equiv {\beta s \over {e_{01}}^\mu} (e+e_{02})^{\mu} -\beta E e $ of the $\sinh$ 
in expression (\ref{dif1}) never vanishes for any value of $e>0$.
If the stress is smaller than the threshold $s^*$, the equation $B=0$ has two solutions 
and the system does not fail. If $s>s^*$, $B$ does not vanish and the 
system fails in finite time. 
If $s$ is close to but larger than $s^*$, $B$ can be expanded as follows around its minimum:
\be
B = c_1 (s-s^*) + c_2 (e-e^*(s))^2~,
\label{aaadev}
\ee
where $e^*(s)$ is the value of the deformation which makes $B$ minimum and $c_1$ and
$c_2$ are two positive constants.
Since $s \to s^*$ from above, $s-s^*$ is small. Also, the lifetime of the system is controlled 
by the domain of deformation giving the smaller strain rate, that is, for deformations $e$ in 
the neighborhood of $e^*(s)$. We can thus expand $\sinh(B) \approx B$ to linear order and still 
obtain the correct scaling of the total lifetime. This leads to approximate equation (\ref{dif1}) by
\be 
{de \over dt} = K~\left[  c_1 (s-s^*) + c_2 (e-e^*(s))^2 \right]~.
 \label{sdmfm}
 \ee
 The change of variable $t \to \tau = t~(s-s^*)^{1/2}$ and $e \to x=e~(s-s^*)^{-1/2}$ 
transforms (\ref {sdmfm}) into
 \be
 {dx \over d\tau} = K~\left[  c_1 + c_2 (x-x^*(s))^2 \right]~.
 \ee
 The lifetime $\tau_c$ is obtained as
 \be
 \tau_c \approx \int_0^{+\infty} dx ~ {1 \over (dx/d\tau)}
 \ee
 which is a pure number independent of $s$. This thus gives $t_c ~(s-s^*)^{1/2} =\tau_c$ and thus
 \be
 t_c \sim 1/(s-s^*)^{1/2}~.
 \label{jgnwlfgqw}
 \ee
This result (\ref{jgnwlfgqw})  holds also for model 1 (Hidalgo et al., 2002; 
Kun et al., 2003), as it is based solely on the
expansion (\ref{aaadev}) and its analogs for model 1, which is
generic. This universality expresses only the topology of a straight
line almost tangent to a smooth curve similar to that shown in 
Figure \ref{stablesol}, such that the distance of the
curve to the straight line is given by an expression such as (\ref{aaadev}). 
For model 2, the dependence (\ref{jgnwlfgqw}) is modified due to the
nonlinear shear-thinning rheology. In this case, expression (\ref{sdmfm})
is changed into
\be
\left({de \over dt}\right)^n \sim \left[ c_1 (s-s^*) + c_2 (e-e^*(s))^2 \right]~.
\label{sdmfm2}
\ee
The change of variable $t \to \tau = t~(s-s^*)^{-{1 \over 2}+{1 \over n}}$ and 
$e \to x=e~(s-s^*)^{-1/2}$ transforms (\ref{sdmfm2}) 
into a non-dimensional equation. This transformation predicts 
$t_c \sim 1/(s-s^*)^{{1\over n} -{1 \over 2}}$ for model 2, which
is verified by our direct numerical integration. 

The dependence of $t_c$ as a function of $s$ for $s \gg s^*$ can also 
be predicted analytically as follows. Since
the time-to-failure $t_c$ is controlled by the primary and secondary regimes
which last significantly longer than the final accelerating tertiary regime,
the dependence of $t_c$ on $s$ is simply obtained from the dependence of the
duration of the primary regime on $s$. This dependence is obtained by rewriting
equation (\ref{pcc}) as
\be
{de \over dy} \approx  e^{ \alpha e }~,
\label{pcc22}
\ee
where
\be
\alpha = {\beta s {e_{02}}^{\mu} \over {e_{01}}^\mu} ~ {\mu \over e_{02}} -\beta E~,
\ee
and
\be
y \equiv t~ {K\over 2} ~e^{ \gamma s }~.
\label{mgmallsd}
\ee
with
\be
\gamma ={\beta  {e_{02}}^{\mu} \over {e_{01}}^\mu}~.
\ee
With the reduced time $y$, expression (\ref{pcc22}) is now dimensionless.
Thus, the duration of this regime and thus $t_c$ scale as 
\be
t_c \sim \exp \left(- \gamma s \right)~.
\label{solutionexpforlarges}
\ee

The predictions (\ref{jgnwlfgqw}) and (\ref{solutionexpforlarges}) for $t_c(s)$ are 
verified in Figure \ref{tcs}.
Contrary to the primary regime, the heterogeneity of the rupture threshold
is an essential ingredient for the observation  of a power-law singularity.
The acceleration toward failure is due to the positive feedback effect of broken RE,
which increases the stress and deformation on the unbroken RE leading
to the global failure of the system. We can observe the same power-law
singularity (but only on  a finite range of times) if we replace the
power-law distribution of rupture threshold by an exponential
distribution $P(e)=1-\exp(-e/e_0)$ with a large enough heterogeneity $e_0$.
As long as the strain $e$ is small compared to $e_0$, we have $P(e)
\approx e/e_0$, therefore the stress $s_1$ in the dashpot is
proportional to $e$ leading to $de/dt \sim 1/(t_c-t)^{p'}$ close to failure. 

\begin{figure}
\includegraphics[width=1\textwidth]{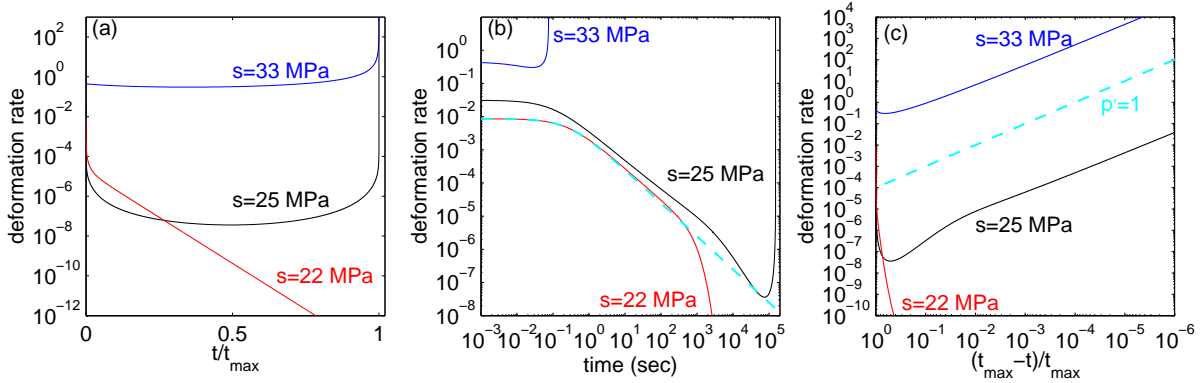}
\caption{\label{num1} Strain rate $de/dt$ estimated by the
numerical solution of eq. (\ref{dif1}) for different values of the
stress $s$, and with parameters $E=20$ GPa, $\mu=1.2$, $e_{01}=0.003$,
$e_{02}=0.015$, $\beta=50$ GPa$^{-1}$ and $K=10^{-5}$ sec$^{-1}$. 
Panel (b) illustrates Andrade's law in the 
primary regime, with exponent $p \approx1$ for $s=22$ MPa 
and $p \approx 0.8$ for  $s=25$ MPa.
Panel (c) illustrates the power law acceleration of the 
strain rate before failure for $s=25$ MPa and $s=33$ MPa with an 
exponent equal to 1 asymptotically but slightly smaller than one
in the crossover further from rupture.
The time $t_{max}$ used to normalize the curves
in (a) and (c) is the rupture time ($de/dt>10^{20}$) for 
$s=25$ MPa and for $s=33$ MPa, and is defined by the time when 
$de/dt$ decreases below $10^{-14}$ (numerical accuracy) for 
$s=22$ MPa. The dashed line in (b) is the approximate solution
(\ref{pc}) of (\ref{dif1}) with $s=22$ MPa for the primary creep 
(power-law with $p=1$ at large $t$).}
\end{figure}

\begin{center}
\begin{figure}
\includegraphics[width=0.5\textwidth]{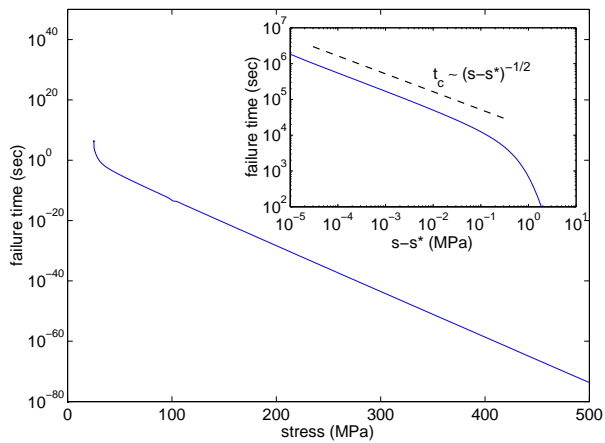}
\caption{\label{tcs} Lifetime $t_c$ as a function of the applied
stress $s$, for the same parameters as in Figure \ref{num1}.
The failure time decreases exponentially with $s$ for $s \gg s^*$
and decreases as a power-law $\sim (s-s^*)^{-1/2}$ for $s\approx s^*$
(see inset).}
\end{figure}
\end{center}

Figure \ref{num1} shows the numerical solution of the differential equation 
(\ref{dif1}) together with the approximate analytical solutions (\ref{pc}) 
in the primary creep and (\ref{mgmss}) close to failure,
for different values of the applied stress $s$.
In contrast with the analytical solution   (\ref{pc}) for the primary creep,
which predicts $p=1$, we observe close to the rupture threshold
$s\approx s^*$ an apparent smaller exponent $p\approx 0.8$ over 2
orders of magnitude in time, which can explain the values of $p<1$
found experimentally (see Table \ref{tabdef}). For a stress $s$ much larger
than $s^*$, the strain rate $de/dt$ accelerates immediately when the
load is applied. For $s<s^*$,  there is an exponential decrease of the
strain rate $de/dt$ after the Andrade regime (see Figure \ref{num1}).
Close to failure for $t\approx t_c$ and $s>s^*$, the numerical integration of 
(\ref{dif1}) shows that ${d e \over dt} \sim 1/(t_c -t)$ (\ref{dif1ss})
is a good approximation very close to failure, but that for $s\approx s^*$ we can
observe a crossover further from failure with an apparent exponent $p'=0.9$.
The numerical integration of (\ref{dif1}) shows that both power-laws in the 
primary and tertiary creep regimes are obtained in this model, with an
apparent exponent $p$ for the primary creep smaller 
than $1$, and with an exponent $p'=1$ for the tertiary regime, except
from a crossover with an apparent exponent $p'$ a little smaller than
one. This crossover with $p'<1$ is not sufficient to explain the observations 
of $p'\approx 0.8$ over 4 orders of magnitude in time $t_c-t$.   

Figure \ref{tcs} shows the failure time $t_c$ as a function of the
applied stress. The failure time has a power-law singularity $\sim
(s-s^*)^{-1/2}$ for $s\approx s^*$, as found previously by 
Hidalgo et al. (2002) and Kun et al. (2003)
for the  first  version of the model (linear dashpot).
The failure time decays exponentially for $s\gg s^*$.
Using a numerical integration of (\ref{dif1}), for different choices
of the parameters, we find that, for $s>s^*$,
the transition time $t_m$ (minima of the strain rate) is always equal to $t_c/2$,
as for the 2 previous models. This result recovers the proportionality of $t_m$ 
and $t_c$ found experimentally, but predicts a shorter duration for the primary creep 
($t_c/2$) than the observations ($t_m \approx 2t_c/3$, see Figure \ref{tmtc}).

This model explains qualitatively the existence of both
the power-law decrease of the strain rate during the primary creep regime and the 
power-law singularity before failure. It also explains the $p$-value of 
the primary creep, which ranges in the model between
0 and 1 as a function of the applied stress. The larger the stress, the smaller
the apparent $p$-value, and the smaller the failure time and the duration of the primary 
creep. If the stress is too large, we do not observe the first primary creep regime
and the strain rate continuously accelerates before failure.
The model thus explains the correlation observed between the $p$-value and
the failure time shown in Figure \ref{cor}e.

We have also studied a more complex model with RE elements that obeys 
Burger's topology, in an effort to explain the value $p'<1$ found experimentally. 
Each RE element is composed of a Kelvin element (spring and dashpot in parallel)
associated in series with a Maxwell element (spring and dashpot in series).
The strain rate in now the sum of the two contributions of the Maxwell
and Kelvin element. We found that adding the Maxwell element in
parallel does not change qualitatively  the behavior of the model.
 If the strain rate of Maxwell's element is similar to the strain rate
of the Kelvin element, we observe close to failure an apparent power-law 
with an exponent  $p'<1$, but for a limited time-range (no more than one order of
magnitude in time), which evolves toward a pure power-law with $p'=1$ close to failure.

\section{Conclusion}

We have performed creep experiments on fiber composites. We have established 
experimentally the existence of power-law relaxations both for the strain
(Andrade's law) and for the acoustic emissions (analog of Omori's law
for earthquakes) in the primary creep regime.
We also observed a power-law singularity both for the strain rate and
for the rate of AE in the tertiary regime culminating in global failure.
Our experiments confirm over large time scales covering up to 4 orders
of magnitude in time
previous announcement of power laws in the tertiary creep regime,
which were established over more limited time scales  (Guarino et al., 2002).

We have shown the correlations between the $p$-exponent
of the primary creep regime and the rupture time  $t_c$. 
We have also shown a strong correlation of $t_c$ with
the time $t_m$ of the minimum strain rate in the secondary creep regime.
These results suggest interesting possibilities for the prediction 
of the rupture time $t_c$.
We have developed a simple model of representative elements, 
made of a spring in parallel with a linear or nonlinear dashpot, 
interacting via democratic load sharing. Our model recovers 
all the observations apart from the exact value of $p'$, which is found 
experimentally to be $\approx 0.8$, smaller than the value $p'\approx1$ found in 
the model. An essential ingredient for the observation of the power-law
singularity of the strain rate before failure is a large heterogeneity
of the rupture thresholds of the representative elements.
We stress that the interactions between the RE elements together
with a large heterogeneity and a simple rheology is sufficient and replaces the need
for complex memory effects. In particular we do not need to invoke the  
integro-differential Schapery long-memory formalism (Cardon et al., 2000; 
S\'egard et al., 2002). 

A natural improvement of the model would be to relax the democratic
load sharing rule a la DFBM, along the work of Kun et al. (2003).
These authors show interesting new effects
in the creep dynamics of Model 1 when introducing realistic elastic interaction
with an elastic Green function decaying as a power law of the distance between
elements. We expect that this improvement will provide
a more realistic value of  $p'$ of the strain in the tertiary creep
regimes. 

We end by on a conceptual note. The model presented here is a macroscopic
deterministic effective description of the experiments. In contrast, the modeling 
strategy of Ciliberto et al. (2001), Politi et al. (2002) and
Saichev and Sornette (2003) emphasizes the interplay between 
microscopic thermal fluctuations and frozen heterogeneity. Qualitatively,
our models  are similar to a deterministic macroscopic Fokker-Planck description 
while the thermal models of Ciliberto et al. (2001), Politi et al. (2002) and
Saichev and Sornette (2003)  are reminiscent 
of stochastic Langevin models. It is well-known in statistical physics that 
Fokker-Planck equations and Langevin equations are exactly equivalent 
for systems at equilibrium and just
constitute two different descriptions of the same processes, and their
correspondence is associated with the general fluctuation-dissipation
theorem. Similarly, the
encompassing of both the Andrade relaxation law in the primary creep regime and of
the time-to-failure power law singularity in the tertiary regime by
our model 3 and by the thermal model as shown by  Saichev and Sornette (2003) 
suggests a deep connection between these two levels of description 
for creep and damage processes. 

\vskip 1cm
{\large \bf Acknowledgments}\\
We acknowledge useful discussions with Francois Sidoroff and
Amen Agbossou. This work is partially supported by
the James S. Mc Donnell Foundation 21st century scientist
award/studying complex system and by the french department of 
research under grant N$^o$ 207.

\end{document}